\documentclass[letterpaper]{article}
\pdfoutput=1
\usepackage{jheppub}
\usepackage{amsmath}
\usepackage{graphicx}
\usepackage[config=altsf]{subfig}
\usepackage{feynmf}
\usepackage{array}
\usepackage{slashed}
\usepackage{afterpage}
\usepackage{appendix}
\usepackage{multirow}

\newcommand{\GeV}{\,\text{GeV}}
\newcommand{\MeV}{\,\text{MeV}}
\newcommand{\keV}{\,\text{keV}}
\newcommand{\cm}{\,\text{cm}}
\newcommand{\km}{\,\text{km}}
\newcommand{\s}{\,\text{s}}
\newcommand{\pc}{\,\text{pc}}
\newcommand{\kpc}{\,\text{kpc}}

\title{Point Sources from  Dissipative Dark Matter}
\author{Prateek Agrawal and}
\author{Lisa Randall}
\affiliation{Department of Physics, Harvard University, Cambridge, MA 02138, USA}

\abstract { If a component of dark matter has dissipative
  interactions, it can cool to  form  compact astrophysical objects
  with higher density than that of conventional cold dark matter
  (sub)haloes.
  Dark matter annihilations might then
  appear as point sources, leading to novel morphology for indirect
  detection. We explore dissipative models where interaction with the
  Standard Model might provide visible signals,  and show how such
  objects might  give rise to the observed excess in gamma rays
arising from the galactic center.  }

\preprint{\today}

\begin{document}
\maketitle
\section{Introduction}

A  vast amount of astrophysical evidence exists for the presence of dark
matter, but its particle physics nature remains entirely unknown. Various
direct and indirect detection experiments that search for  small
non-gravitational interactions between
the dark and Standard Model sectors  hope to find this dark
component of our energy density.

If the Standard Model (SM) is a guide, the dark sector may contain multiple
species, with small components possessing stronger  interactions,
amongst itself and/or with the SM. Even if this component
is subdominant -- so that it does not impact gravitational
measurements of dark matter so far -- it might generate the first
signals we discover.

One such class of models is where such a  dark matter fraction  has
dissipative dynamics~\cite{Fan:2013yva,Fan:2013tia}. In analogy with
baryonic matter, this could lead
to the formation of interesting astrophysical structures, without
disrupting qualitative features of
cold dark matter halo formation. If this component also interacts
non-gravitationally with the Standard Model, the first signals of
dark matter might have unexpected
properties~\cite{CyrRacine:2012fz,McCullough:2013jma,Fan:2013bea,Fischler:2014jda,2014PhRvL.112p1301R,Randall:2014kta,Reece:2015lch,Kramer:2016dew,Kramer:2016dqu,Agrawal:2017rvu}.

In this paper, we consider the possibility that dark matter forms
dense clouds that are sites of enhanced annihilation. These novel
astrophysical structures -- such as a dark disk or dense compact halo
objects -- will give rise to distinctive morphology for indirect
detection.  If small enough, these clouds would appear as point
sources in indirect detection experiments. We explore models with both
dissipative interactions and couplings to the SM that can give rise
to visible signatures from dark compact objects, with the specific
signal depending on the particular nature of this coupling.  We will
see that one of the most interesting aspects of such models is that
the dissipative dark matter can be a negligible fraction of the total
dark matter composition, yet still give rise to observable effects.

An important signal is the excess of gamma rays observed from towards
the galactic
center~\cite{Goodenough:2009gk,Hooper:2010mq,Daylan:2014rsa,Calore:2014xka,TheFermi-LAT:2015kwa}, with some statistical evidence supporting this
excess as arising from point
sources~\cite{Lee:2015fea,Bartels:2015aea,Fermi-LAT:2017yoi}, 
though this may be due to substructure in
background that is not modeled in the range of diffuse background
models considered (see also~\cite{Horiuchi:2016zwu}).  The  more
conventional assumption
for sources that would  produce the observed point-like spectrum could
be a new population of millisecond pulsars (MSPs), with luminosity
just below the Fermi point source detection threshold.  Direct
evidence for such a population has not yet been observed (see also
\cite{Brandt:2015ula,Hooper:2016rap,Fermi-LAT:2017yoi}).  If we are to
definitely
establish the origin, it is worthwhile to consider less conventional
potential sources as well.  An alternative admittedly more speculative
possibility that is also worth investigating  is that the excess is
associated with dark point sources. 
In this paper we consider pointlike sources that arise from
dissipative dynamics in the dark sector.
In this scenario the dark objects cool analogous to the objects in the
SM sector. In our case we only have a $U(1)$ interaction that leads to
cooling, and we study the compact objects that would result from this
interaction. An alternative is ultra-compact minihaloes
(UCMHs)~\cite{Berezinsky:2003vn,Ricotti:2009bs,Scott:2009tu,Josan:2010vn,Erickcek:2011us,Bringmann:2011ut,Berezinsky:2013fxa,Clark:2015sha,Aslanyan:2015hmi,Clark:2016pgn}
and we discuss ways to distinguish
them at the end.

This paper is organized as follows. We start in
section~\ref{sec:compact-objects} explaining how dissipative dark
matter can give rise to compact objects and present a model-dependent
estimate for their expected sizes.  In section~\ref{sec:model} we
introduce a simple dissipative dark matter model and
specific portal couplings to the
Standard Model. We analyze the resulting phenomenology in
section~\ref{sec:phenom}. Section~\ref{sec:GCE} shows how such a model
could account for the Galactic Center excess, and we conclude in
section~\ref{sec:conclusion}.

\section{Compact objects from dissipative dark matter}
\label{sec:compact-objects}

If there is a component of dark matter that has dissipative dynamics,
it can cool efficiently with the potential creation of compact
objects. The sizes and distribution of such objects depends on
nonlinear dynamics, and a careful prediction would require modeling and
simulations. 
However, simple estimates demonstrate that compact
objects, with size determined by fraction of dark matter in the
charged dark component, are a viable possibility. 

We assume the dissipative component leads to formation of an unstable
disk which then fragments via Toomre instabilities to form compact
objects.  An alternative possibility is that compact objects form in
smaller subhaloes and might survive as subhaloes merge to form our
galaxy. This latter scenario could potentially lead to compact objects
with a distribution different from the one we now consider but is
worth independent consideration.

When formed from disk fragmentation, the typical size of the objects
can be estimated using a simple
stability analysis, which we perform below. 
For simplicity, we assume that the clumps are uniform so that we can
characterize  parameterize  the distribution and mass functions
with a single characteristic size and mass,
$\{M,R\}$, dictated by the stability analysis.  
More generally, one could parameterize the number of
objects of mass $M$ by the mass function, $dN/dM$ and assume a
nonuniform density even within the objects. 

We will consider a simple dissipative model with a weak-scale particle
$X$ and a light particle $C$, both charged under a dark $U(1)$ with
coupling strength $\alpha_D$. This model is discussed in further detail in
section~\ref{sec:model}.

\subsection{Dark Disk Fragmentation and the size of clouds}
\label{sec:fragments}

The stability
condition for the dark disk in the presence of various gas and stellar
components  can be written using the dispersion  relation for a
density perturbation
\cite{
Romeo1992,Rafikov2001,
binney2008galactic,Randall:2014kta,Shaviv:2016umn},
\begin{align}
2\pi G k
  \sum_{\mathrm{collisional} }\frac{\Sigma_i}{\kappa^2 + k^2
  c_i^2 - \omega^2}
  +
2\pi G k
  \sum_{\mathrm{collisionless} } \frac{\Sigma_i
    \mathcal{F}(\omega/\kappa,k^2 \sigma_i^2/\kappa^2)}{\kappa^2 - \omega^2}
    &=
    1
    \label{eq:dispersion}
\end{align}
The gas and the dissipative component of dark matter are collisional, while
the various population of stars are collisionless. 
The epicyclic frequency $\kappa=36\km \s^{-1} \kpc^{-1}$, and $k$ denotes
the wavenumber of the perturbation.
$\Sigma_i$ is the column density of component $i$, and
$c_i$($\sigma_i$)
is the sound speed (velocity dispersion) of the collisional
(collisionless) component. The function $\mathcal{F}$ is 
\begin{align}
  \mathcal{F}(s,\chi)
  &=
  \frac{2(1-s^2)e^{-\chi}}{\chi}
  \sum_{n=1}^{\infty} \frac{I_n(\chi)}{1-s^2/n^2}
  \,.
\end{align}
$I_n$ is the Bessel function of order $n$.
The perturbation is unstable when $\omega^2 < 0$. Since the l.h.s
in equation~\ref{eq:dispersion} is a monotonically increasing function
of $\omega$ when $\omega<\kappa$, the
instability criteria
can be restated as the condition that the l.h.s is greater than 1 at
$\omega=0$,
\begin{align}
2\pi G k
  \sum_{\mathrm{collisional} }\frac{\Sigma_i}{\kappa^2 + k^2
  c_i^2}
  +
2\pi G k
  \sum_{\mathrm{collisionless} } \frac{\Sigma_i
    \mathcal{F}(0,k^2 \sigma_i^2/\kappa^2)}{\kappa^2}
    &>
    1
    \,.
\end{align}

The sound speed for the dark disk can be estimated by the
temperature dissipative dark matter cools down to, which is
roughly the temperature at which recombination occurs and rapid
cooling ceases: $T\sim
\frac{1}{40} m_C \alpha_D^2$,
\begin{align}
  c_{XC}
  \sim
  \frac{\alpha_D}{\sqrt{40}}
  \sqrt{\frac{m_C}{m_X}}
  \label{eq:dmvel}
\end{align}

\begin{figure}[htp] \centering
  \includegraphics[width=0.65\textwidth]{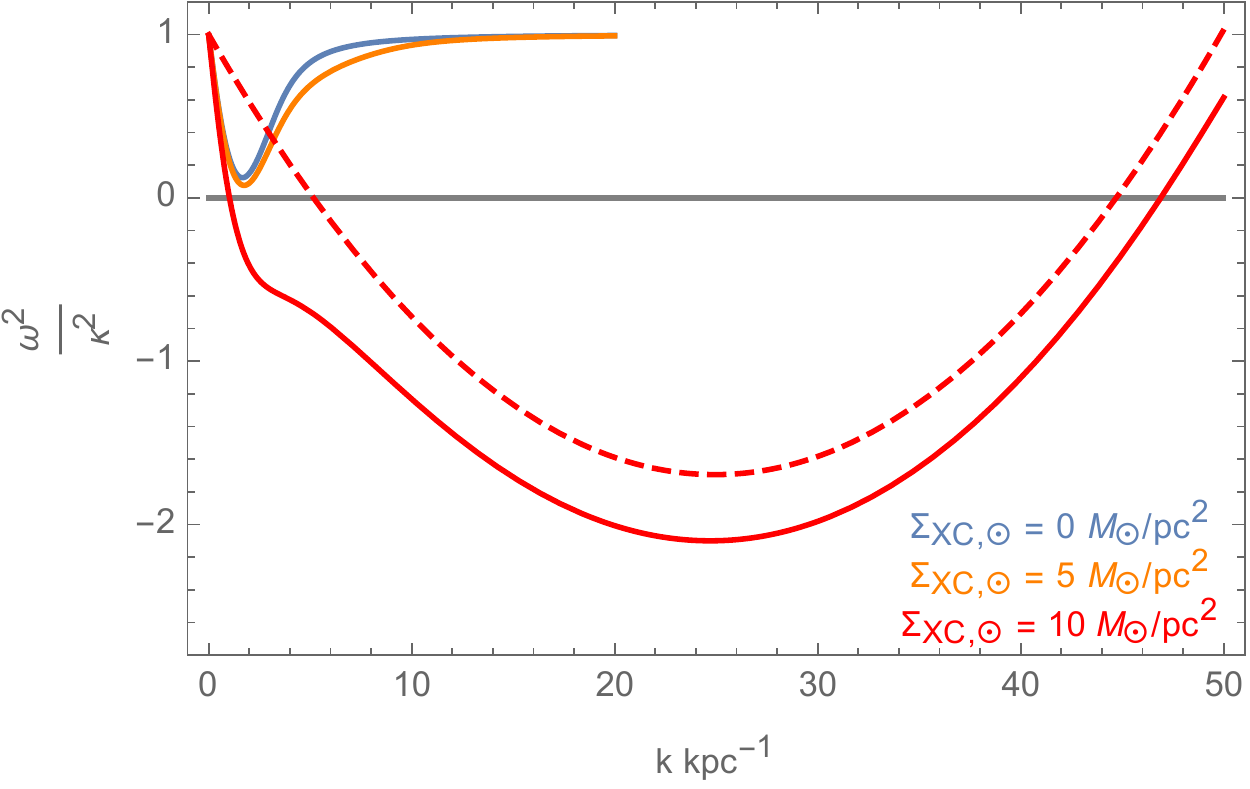} 
  \caption{Instability ($\omega^2<0$) in the galactic disk at the
    solar system as a function of the wavenumber of the perturbation
    for different column densities of the dark disk. We also show the
    instability curve obtained by considering the dark disk in
    isolation (dashed). The $XC$ dark matter sound speed was taken as 2.5 km/s
    (corresponding to $\alpha_D = 0.01, m_C=0.2\MeV, m_X=25\GeV$). Values
    for star and gas components were taken from McKee et
    al~\cite{2015ApJ...814...13M}.
   \label{fig:toomrenew}}
\end{figure}

We show the stability condition in figure~\ref{fig:toomrenew}. Note
that in the Milky Way, galaxy stars and gas are  relevant
to stability only on larger length scales. For our choice of parameters,
the dark disk induces instability at much shorter length scales, where
the contribution of
stars and gas to the dispersion relation can be neglected. It
therefore suffices here to to consider the
stability of the dark disk by itself and
find
a simple expression for the wavenumber of unstable modes $k_- < k
<k_+$,
\begin{align}
  k_\pm
  &=
  \frac{\pi G \Sigma_{XC}}{c_{XC}^2}
  \pm
  \sqrt{
    \left(\frac{\pi G \Sigma_{XC}}{c_{XC}^2}\right)^2
    -\frac{\kappa^2}{c_{XC}^2}
  }
  \,.
\end{align}
The fastest growing mode corresponds to $k = 
  {\pi G \Sigma_{XC}}/{c_{XC}^2}
  $.

We assume 
the dark disk surface density is
\begin{align}
  \Sigma_{XC}(r)
  &=
  \Sigma_{XC,\odot}
  e^{-(r-r_\odot)/r_d}
\end{align}
where $r_d\simeq 3$ kpc is the assumed scale radius of the dark disk, and 
$\Sigma_{XC,\odot}$ is the 
local dark disk surface density which can be as large as $\sim10
M_{\odot}/\pc^2$~\cite{Kramer:2016dqu}. Therefore the wavenumber
varies as a function of distance from the
center since the density and hence the stability criterion vary.

\begin{figure}[tp]
  \centering
  \includegraphics[width=0.45\textwidth]{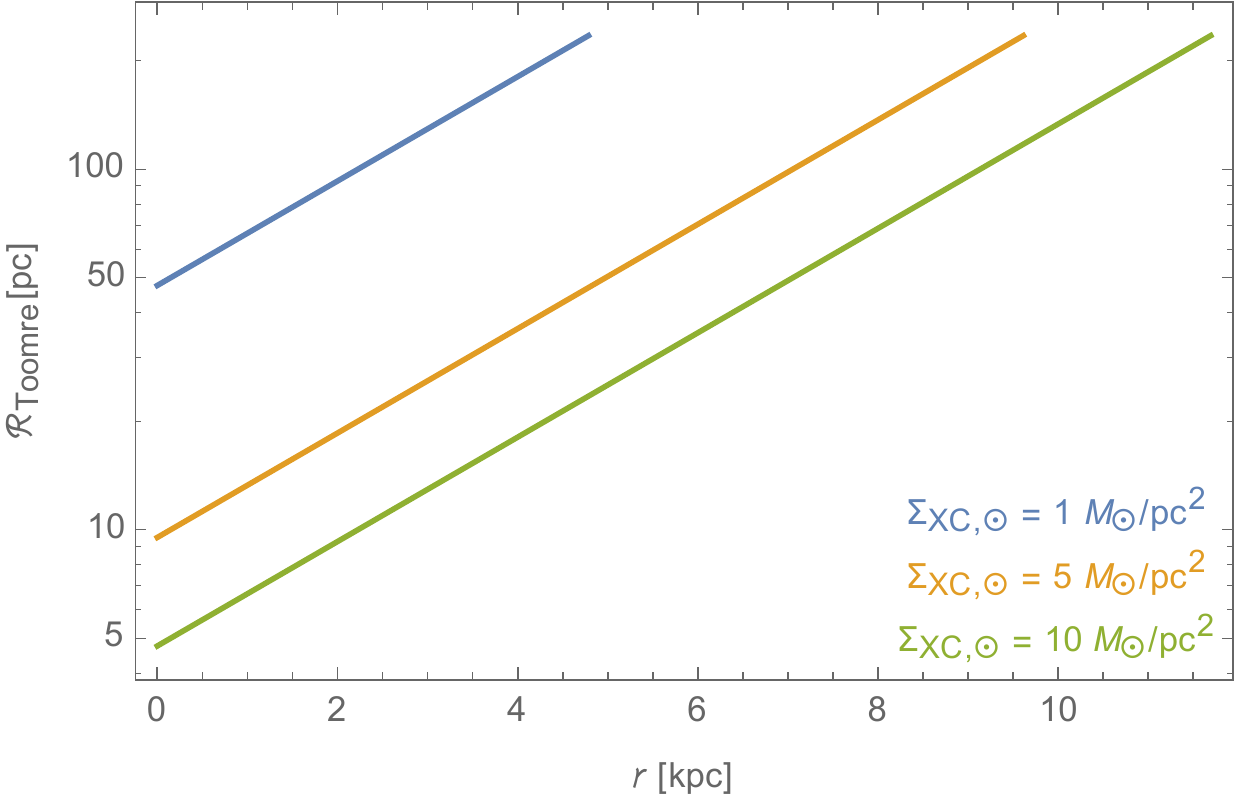}
  \quad
  \includegraphics[width=0.45\textwidth]{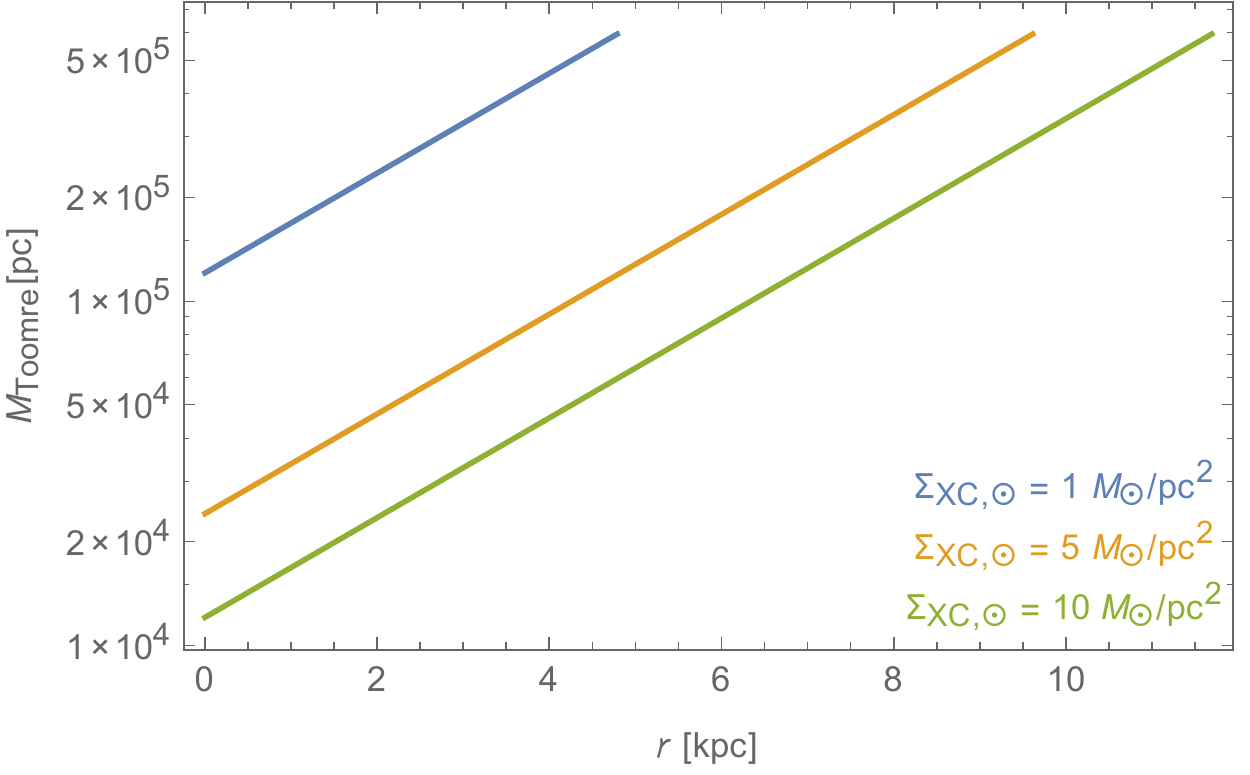}
  \caption{The initial size and mass of the fastest growing Toomre
    instability as a function of distance from the Galactic Center.
    The parameters here are $\alpha_D=0.01$, $m_C = 0.2$
    MeV, $m_X=25 \GeV$, and $r_d=3$
  kpc.}
  \label{fig:toomre}
\end{figure}

The dark compact objects that will be formed then have mass approximately
$M=\pi \Sigma_{XC} \mathcal{R}_{Toomre}^2$, where
$\mathcal{R}_{Toomre}$
is the wavelength of the fastest growing mode.
In figure~\ref{fig:toomre} we show the size and mass of this
mode as a function of the distance from the
Galactic Center. We see that in this scenario we expect  objects of mass 
$10^4 M_\odot$, with
a radius $\sim 10$ pc in the inner part of the galaxy. Of course, this is
a very crude estimate of the initial size, and this object might
undergo further collapse, tidal disruption,  evaporation or accretion.

We note that the spatial distribution of the objects would require
more detailed simulations.  Clumps might be dominantly clustered in
the inner galaxy, spread through the disk, or distributed throughout
the dark matter halo.  We will assume that the dark matter also forms
a disk + bulge system, with the disk subject to fragmentation.  By
analogy with the bulge, this would lead to a population of dark
compact objects at the center of the galaxy in a roughly spherical
distribution. In principle compact objects can form throughout the
bulk as well, though the answer would depend on whether such objects
fall to the center and shock heating repeats itself. For now we assume
compact objects are concentrated in the center and the disk, and ask
what the consequences for indirect detection might be. We will focus
on the objects in the inner galaxy, and not consider
other potential visible signals from the plane of the disk or the
halo
which would require detailed simulations.

\section{Dissipative Dark Matter Models}
\label{sec:model}

Dark matter with dissipative dynamics was considered in
Ref. \cite{Fan:2013yva,Fan:2013tia}. In this section we expand the
discussion to include the possibility that in addition to dissipative
self-interactions there are also interactions with the Standard Model.
In this case we can have dissipative interactions that lead to  the
formation of compact objects and these in turn can   potentially give
rise to visible signatures.

\subsection{A simple dissipative dark matter model}
The model consists of
 a heavy dark proton $X$ and a lighter dark electron $C$, both
charged under a dark U(1) gauge group, as
in~\cite{Fan:2013yva,Fan:2013tia,Cline:2012is}.
\begin{align}
  \mathcal{L}
  &=
  -\frac14 V_{\mu\nu} V^{\mu\nu}
  +e J^\mu_{EM} A_\mu
  +g V_\mu
\left(
\bar{X} \gamma^\mu X + \bar{C}\gamma^\mu C
\right)
  \label{eq:ddm}
\end{align}

A sufficiently large relic population of a light enough $C$ would 
lead to efficient
cooling in dark matter haloes. Being light, $C$s would annihilate
efficiently, and
survive  only if produced through non-thermal processes -- 
such as a matter-antimatter asymmetry in the population.  We consider
models in which
there is both a symmetric relic abundance of $X$ and
$\bar{X}$ particles, as well as an asymmetric population of $C$ and $X$.
Such a symmetric population would occur naturally for dark matter
masses comparable to the weak scale, just as for the more usual WIMPs.
By charge neutrality $n_C - n_X + n_{\bar{X}} = 0$. We define
the fraction of
$XC$ dark matter in the asymmetric component by
\begin{align}
  f_{(XC)}
  &\equiv
  \frac{\rho_C + \rho_X - \rho_{\bar{X}}}
  {\rho_C + \rho_X + \rho_{\bar{X}}}
  \approx
  \frac{n_X - n_{\bar{X}}}{n_X + n_{\bar{X}}}
  \label{eq:asymm}
\end{align}
We also define the fraction of dark matter density in the charged
system,
\begin{align}
  f
  &=
  \frac{\Omega_{X}}{\Omega_{DM}}
\end{align}
with $\Omega_X$ the energy total density in $X, \bar{X}$ and $C$. The rest of
the dark matter is assumed to be made up of a distinct cold dark
matter (CDM) component.

To generate visible indirect signatures, the model must  also include
``portal'' interactions with the SM.
 A simple option is a kinetic mixing portal, in which the dark photon
 kinetically
mixes with the SM photon. However, constraints on millicharged
particles coupled with constraints on the survival of the compact
objects necessitate an additional dark gauge $Z'$, which can be either
massless or massive-- resulting in different signatures that we
outline below.

\subsection{Portal Models}
\label{ssec:portal}

We first pursue the simplest possibility for interactions between the dark and Standard Model sectors and see why that is not so
promising for current indirect detection measurements. Dark matter can
interact with the SM  through
kinetic mixing of the dark photon with the SM
photon~\cite{Okun:1982xi,Holdom:1985ag},
\begin{align}
  \mathcal{L}
  &=
  -\frac{\epsilon}{2} \frac{e}{g_D} F_{\mu\nu} V^{\mu\nu}
  \, .
  \label{eq:kin-mix}
\end{align}
When the dark photon mass is zero, it is convenient to
redefine the dark photon
\begin{align}
  V_\mu \to V_\mu - \epsilon \frac{e}{g_D} A_\mu
\end{align}
After rescaling $V$ to
obtain a canonical kinetic term, this yields the Lagrangian (to
leading order in $\epsilon$)
\begin{align}
  \mathcal{L}
  &=
  -\frac14 F_{\mu\nu} F^{\mu\nu}
  -\frac14 V_{\mu\nu} V^{\mu\nu}
  +e J^\mu_{EM} A_\mu
  + g_D V_\mu( \bar{X} \gamma^\mu X + \bar{C}\gamma^\mu C)
  + e \epsilon A_\mu ( \bar{X} \gamma^\mu X + \bar{C}\gamma^\mu C)
\end{align}
There is no direct coupling of $V$ to
matter in this basis, but the dark matter particles pick up a
millicharge $e \epsilon$ under the SM photon~\cite{Holdom:1985ag}. 

\begin{figure}[t]
  \centering
  \includegraphics[width=0.65\textwidth]{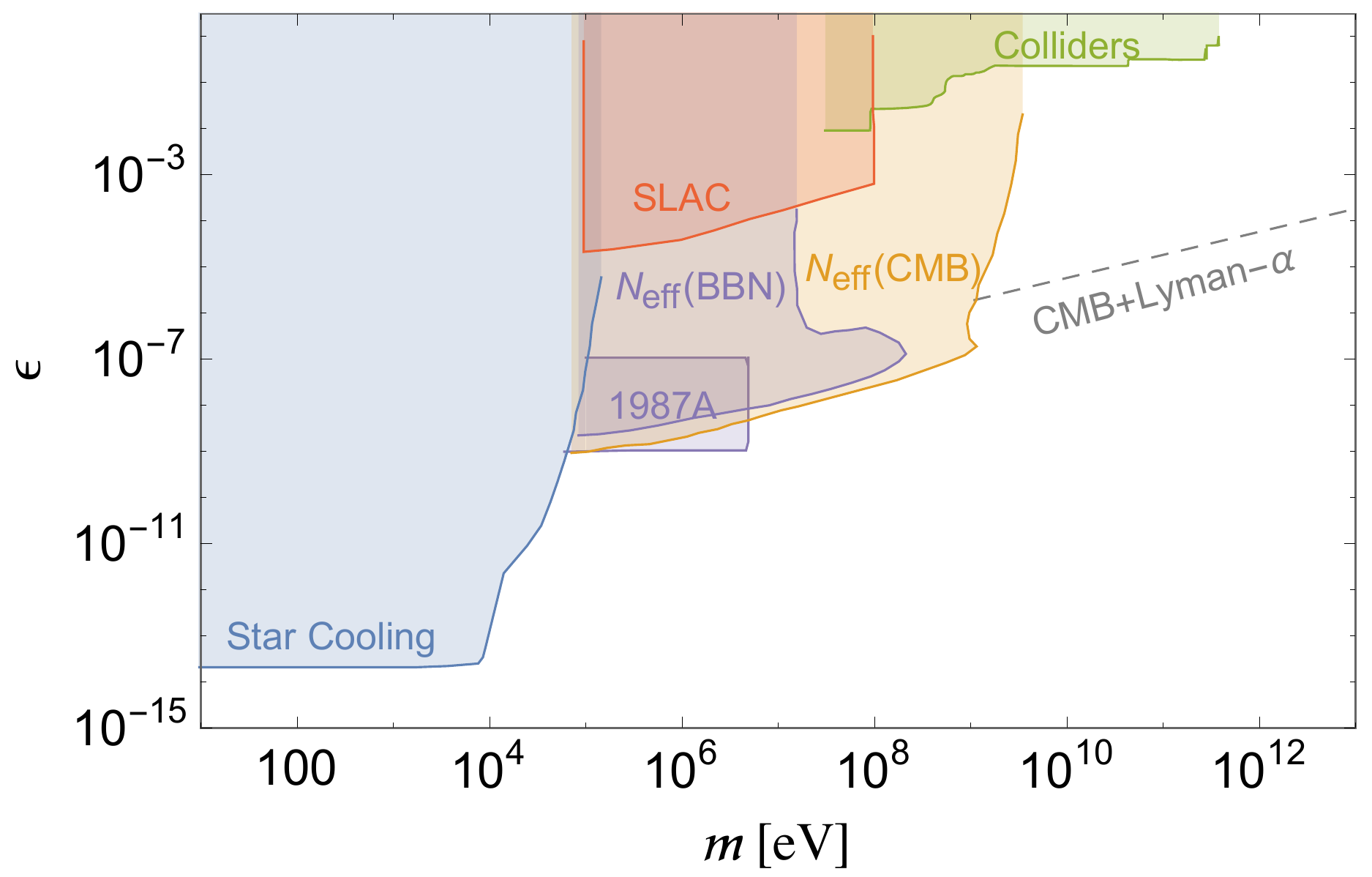}
  \caption{Constraints on millicharged particles from star and
    supernova
    cooling~\cite{Vogel:2013raa,Vinyoles:2015khy}, $N_\mathrm{eff}$
    during BBN and
    CMB~\cite{Brust:2013xpv,Vogel:2013raa}, collider
    experiments~\cite{Davidson:1991si,Prinz:1998ua,
    Davidson:2000hf,Jaeckel:2012yz}. The gray dashed line shows
    the constraint from CMB
    anisotropy and Lyman-$\alpha$
    measurements~\cite{Dubovsky:2003yn,Dolgov:2013una,Dvorkin:2013cea}
    assuming that
    the MCP makes up all of the dark matter.
}
  \label{fig:mcp}
\end{figure}

Models with millicharged particles (MCPs) have rich phenomenology, and the
millicharge has many constraints depending on the mass of the
particle~\cite{Davidson:1991si,
Davidson:2000hf,
Dubovsky:2003yn,
McDermott:2010pa,
Wilkinson:2013kia,
Dolgov:2013una,
Brust:2013xpv,
Dvorkin:2013cea,
Vogel:2013raa,
Vinyoles:2015khy,
Munoz:2015bca,
Ali-Haimoud:2015pwa,
Kadota:2016tqq,
Kamada:2016qjo}.
We summarize the constraints in figure~\ref{fig:mcp}.

For very light particles, $m_C < 10\keV$, star cooling
constraints restrict $\epsilon <
10^{-14}$~\cite{Vogel:2013raa,Vinyoles:2015khy}.
In the range of $C$ masses
where cooling occurs efficiently, $m_C \sim$ MeV,  very
stringent bounds on $\epsilon$ are needed to prevent
dark photons from contributing $N_\mathrm{eff}\sim1$ to the cosmic
microwave background (CMB) and during
Big Bang nucleosynthesis (BBN)~\cite{Brust:2013xpv,Vogel:2013raa}.
The bound relies on thermal coupling and corresponds to when the
sectors thermally decouple. The
dark sector is coupled to the photon bath through the process $e^+ e^-
\to C \bar{C}$. The rate for this process relative to
Hubble is,
\begin{align}
 \frac{ n_e \langle \sigma v \rangle}{H}
  &\sim
  \frac{g T^3}{\pi^2}
  \frac{\pi \alpha ^2 \epsilon^2}{T^2}
  \frac{M_{pl}}{T^2}
\end{align}
Thus, the condition that the dark sector not get into thermal contact
with the SM at temperature $T=m_C$,
\begin{align}
  \epsilon
  \lesssim 
  \left( \frac{m_C}{\alpha^2M_{pl}}\right)^\frac12
  \sim
  10^{-9}
  \left( \frac{ m_C}{1\ \mathrm{MeV}}\right)^\frac12
\end{align}
This is an extremely severe bound and forces $\epsilon$
to be too small to produce observable indirect detection
signals. Interesting signals arise only in the presence of additional
interactions, as we now propose.

\subsubsection{A massless $Z'$ portal}
 
\label{sec:masslesszp}
\begin{table}[ht]
\begin{align}
  \begin{array}{crr}
     \hline
     \qquad\qquad &  U(1)_{D} & \quad U(1)_{Z'}\\
     \hline
     \hline
    X    & 1 & 1  \\
    X^c  & -1 & -1 \\
    C    & -1 &  0\\
    C^c  & 1 &  0\\
    Y    & -1 & 1 \\
    Y^c  & 1 & -1 \\
    \hline
  \end{array}
\end{align}
\caption{Particle content for the massless $Z'$ portal model.
\label{tab:MasslessZprimeContent}
}
\end{table}
The kinetic mixing of the dark photon with the SM photon leads
to a millicharge for the light particle $C$, which is also severely
constrained.  However, an additional massless gauge boson,
$Z'$, which couples to $X$ but not to $C$, can have a moderate
kinetic mixing with the hypercharge gauge boson and produce observable
signals at direct and indirect detection experiments, or at colliders.

As an example, we consider a model with 3 $U(1)$ gauge symmetries --
hypercharge $U(1)_Y$, the dark
photon $U(1)_D$ and the $Z'$ portal $U(1)_{Z'}$. The SM
particles are not charged under the $U(1)_D$ or the $U(1)_{Z'}$, and
the dark sector particles are not charged under the SM gauge group. We
assume that there is no kinetic mixing between the $U(1)_D$ and either
$U(1)$, only between $U(1)_{Z'}$ and $U(1)_Y$,
\begin{align}
  \mathcal{L}
  &=
  -\frac{\epsilon}{2} \frac{e}{g_{Z'}} Z'_{\mu\nu} F^{\mu\nu}
\end{align}

We summarize the particle content of the model 
in table~\ref{tab:MasslessZprimeContent}. The presence of only vectorlike
fermions beyond the SM ensures that there are no anomalies. If we want
to have a cosmological population of $XC$ atoms (which have a net
charge under the
$U(1)_{Z'}$), we need a corresponding population of $Y$ particles to
ensure that the universe is charge neutral 
under $U(1)_{Z'}$.
We note that for this particular choice of charges, there is no
kinetic mixing induced by $X,Y,C$ loops
between the $U(1)_D$ and the $U(1)_{Z'}$. For
simplicity we assume below that the $Y$ particle has roughly the same mass
as $X$, $m_Y \sim m_X$.

The dominant constraints on this scenario come from cosmological and
astrophysical observations (for proposals to cover this parameter
space in collider experiments
see~\cite{Haas:2014dda,Gabrielli:2016rhy}).
There are bounds on this scenario from CMB and
Lyman-$\alpha$
measurements~\cite{McDermott:2010pa,Dvorkin:2013cea,
Dolgov:2013una}.
The bound on kinetic mixing assuming $X$ makes up all of dark matter is
\begin{align}
  \epsilon
  \lesssim
  1.8\times 10^{-6}
  \left(\frac{m_X}{\GeV}\right)^{1/2}
  \label{eq:kin-mix-massless}
\end{align} 
in order for $X$s to be kinetically decoupled from baryons during
recombination. For larger values of the mixing, $X$ and $Y$ can only
make up small fraction, of the total dark matter~\cite{Dolgov:2013una}
, $\Omega_X<0.001$ (where $\Omega_X$ is the energy density of the free
$X,Y$ particles). There are also potentially strong constraints
($\epsilon\lesssim 10^{-14} (m_X / \GeV)$) from the measurement of
shapes of cluster haloes, which would be affected due to the cluster
magnetic force on the MCP dark matter~\cite{Kadota:2016tqq}. However,
this constraint is likely to be dramatically weakened if the MCP makes
up only a fraction of the dark matter density.

A promising signal in this model arises from the annihilation of dark
matter to SM photons through $X$ millicharge~\cite{Aisati:2014nda}.
The annihilation cross
section for $X \bar{X}
\to \gamma Z',\gamma\,\gamma_D$ is, 
\begin{align}
\langle \sigma v\rangle_{X\bar{X}\to \gamma Z'/\gamma_D}
  &=
  \frac{\pi \epsilon^2 \alpha (\alpha_{Z'}+\alpha_D)}{m_X^2}
\end{align}
There is also continuum gamma-ray emission originating from $X\bar{X}$
annihilations to SM fermions through an $s$-channel photon. However,
for comparable mass and couplings, as is expected in a dissipative
scenario, the relative rate of
the line emission and continuum emission are comparable:
\begin{align}
  \frac{\langle \sigma v\rangle_{X\bar{X}\to \gamma Z'/\gamma_D}}
  {\langle \sigma v\rangle_{X\bar{X}\to f \bar{f}}}
  &=
  \frac{\alpha_{Z'}+\alpha_D}{\alpha}
\end{align}
The FERMI sensitivity for the line
signal~\cite{Ackermann:2015lka} 
is about 3 orders of magnitude stronger than the continuum
emission sensitivity. This implies that for
$\alpha_{Z'}+\alpha_D>10^{-5}$, the line constitutes a more promising
avenue for detection of this model.

A search for a line spectrum from a point source will be interesting
to carry out since it has the cleanest spectrum and morphology in
indirect searches. A spectral analysis of unassociated point sources
will be relatively simple, but a stacked analysis or a statistical
analysis along the lines of~\cite{Lee:2014mza,Lee:2015fea} would
likely have a higher sensitivity. Note that even in presence of a
small mixing as in equation~\ref{eq:kin-mix-massless}, it can be possible
to get an observable signal from point sources.

\subsubsection{A massive $Z'$ portal}
We now consider a
massive $Z'$, which in some respects
most closely mimics the gauge structure in the Standard Model. In this
case the dominant signal is continuum photons. As above, $C$ does not
couple to $Z'$. 
In this
case, we do not require the presence of the additional fermion $Y$.
This model is similar in some respects to that considered
in~\cite{Izaguirre:2015eya}.

We can write the leading interactions of the $Z'$ as,
\begin{align}
  \mathcal{L}
  &=
  -\frac{\epsilon}{2} Z'_{\mu\nu} F^{\mu\nu}
  + Z'_\mu  g_{Z} \bar{X} \gamma^\mu X
  +m_{Z'}^2 Z'_\mu Z'^\mu
\end{align}
We  work in the basis where only the visible matter coupling to the
$Z'$ boson is through kinetic mixing, and the dark
matter does not
pick up any charge under the usual photon. This interaction can 
produce signals at direct or indirect detection experiments, as
well as collider signals. Our focus is the
indirect detection signal of an excess in gamma rays from the Galactic
Center. In that respect, the annihilation to $Z'$  pairs mimics the
annihilation to the $b$-quarks in the models of Ref.
\cite{Daylan:2014rsa,Berlin:2014tja,Agrawal:2014una,Izaguirre:2014vva},  which
is seen to give a good fit.

\begin{figure}[t]
  \centering
  \includegraphics[width=0.65\textwidth]{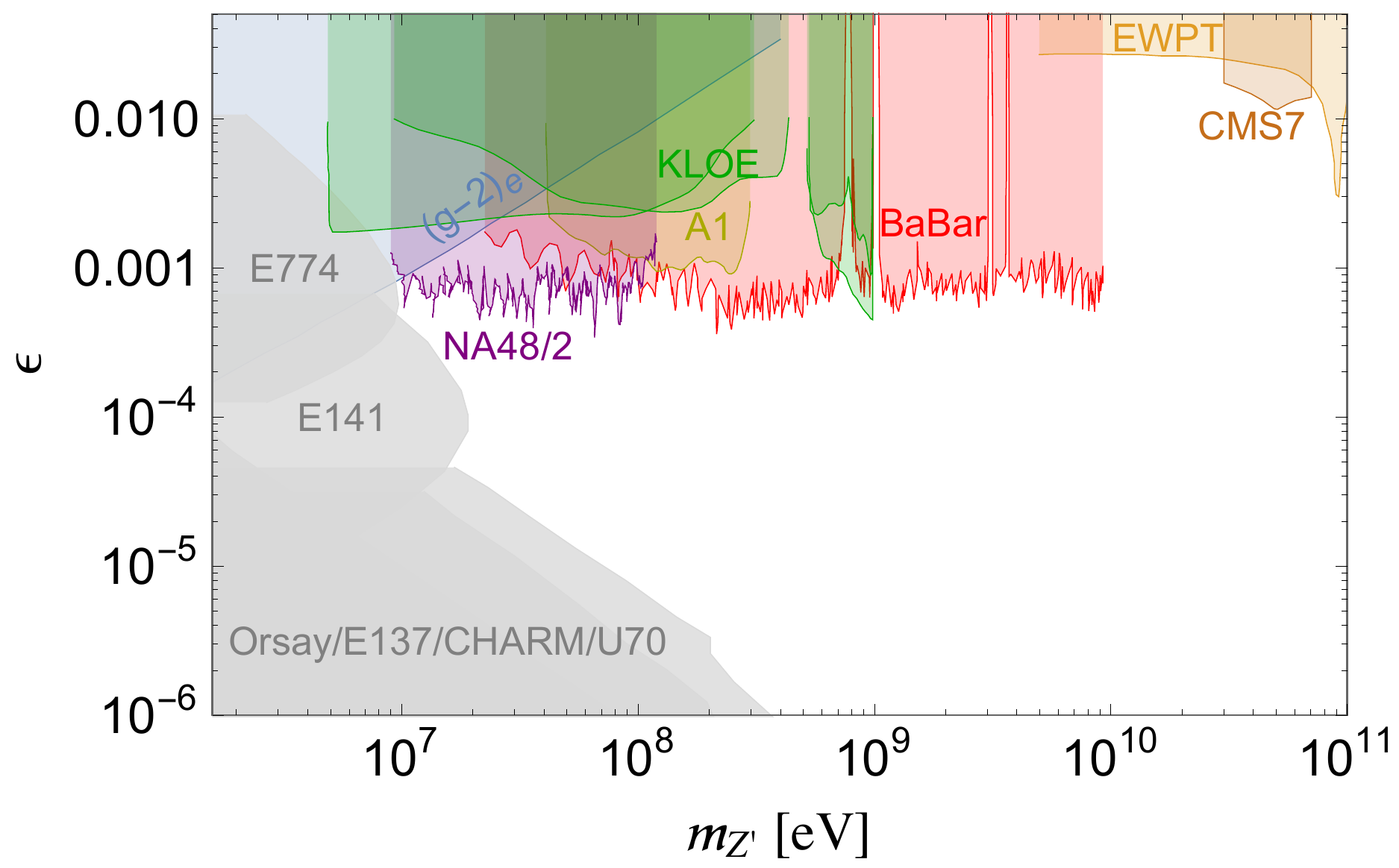}
  \caption{Constraints on the kinetic mixing parameter for a 
  massive $Z'$ portal as a function of the $Z'$ mass. The constraints
  shown are from
  beam-dumps~\cite{Riordan:1987aw,Bjorken:1988as,Bross:1989mp,Blumlein:2011mv,Andreas:2012mt},
  supernovae~\cite{Dent:2012mx,Kazanas:2014mca}, BaBar~\cite{Lees:2014xha},
  A1~\cite{Merkel:2011ze},
  KLOE~\cite{Archilli:2011zc,Babusci:2012cr,Babusci:2014sta,Anastasi:2016ktq},
  anomalous magnetic moment of
  the electron~\cite{Endo:2012hp,Baron:2013eja}, the NA48/2 experiment at CERN~\cite{Batley:2015lha}, 
  electroweak precision measurements~\cite{Curtin:2014cca} and
  Drell-Yan production at the LHC~\cite{Hoenig:2014dsa}.}
  \label{fig:zprime-lim}
\end{figure}

We show the collider limits on the massive $Z'$ portal in
figure~\ref{fig:zprime-lim} (also see~\cite{Alexander:2016aln} for a
detailed review of constraints and future prospects, and for further
references). A number of future experiments/analyses aim to cover new
parameter space for this model~\cite{Echenard:2014lma,Battaglieri:2014hga,
Balewski:2013oza,Abrahamyan:2011gv,Essig:2010xa,Bjorken:2009mm,Anelli:2015pba}.
In the next section we study in detail the phenomenology of
dissipative dark matter with the massive $Z'$ portal.
This model can potentially explain
the excess gamma ray photons observed from the galactic center, which
we return to in section~\ref{sec:GCE}.

\section{Phenomenology}
\label{sec:phenom}
\subsection{Relic abundance} 

A crucial ingredient for dissipative
dynamics is the presence of the light $C$ particles. This requires an
asymmetry in $C$, since the symmetric population annihilates
efficiently. On the other hand, the indirect detection signal from
$X\bar{X}$ annihilation requires a symmetric component to survive as
well. Thus, we are led to consider a freezeout where both a
symmetric as well as an
asymmetric component survives.  

An alternative where the halo also consisted of charged dark matter
was considered in~\cite{Agrawal:2016quu,Agrawal:2017rvu}. Here 
we restrict to the possibility that only
a fraction of the total dark matter is in $X,\bar{X},C$
particles.

A detailed calculation of the relic abundance in a very similar model
was presented in~\cite{Agrawal:2017rvu} (see also 
\cite{Feng:2008mu,Graesser:2011wi,Baldes:2017gzw}), so here we just
recall the important results. 
The key differences from the standard relic abundance
calculation are the presence of an asymmetric component and
a relative temperature between the dark and SM sectors,
$\xi=T_D/T$.
If the interactions between $X$ and the SM are small, the dominant
annihilation channels for $X$ will be in the dark sector, 
$\gamma_D \gamma_D/ Z'
Z'/\gamma_DZ', C \bar{C}$.
We define the ratio of the population of $\bar{X}$ and $X$ as $r$,
\begin{align}
  r
  &=
  \frac{Y_{\bar{X}}}{Y_X}
\end{align} 
such that the value of the ratio today, $r_\infty$,
characterizes how
much of the symmetric and asymmetric components survive.
We define the asymmetry,
\begin{align}
  \eta = Y_X - Y_{\bar{X}}\, = Y_C - Y_{\bar{C}} .
\end{align} 
$\eta$ is a constant that is set at early times, and we treat it as a
free input parameter. 

The Boltzmann equation in terms of $r$ and $x = m_X/T$ is,
\begin{align}
  \frac{dr}{dx}
  &=
  -\frac{\lambda(x) \eta}{x^2}
  \left[ r - r_{eq}(x)
  \left(\frac{1-r}
  {1-r_{eq}(x)}
  \right)^2
\right]
\label{eq:boltzmannr}
\end{align}
We have defined $\lambda$ and $r_{eq}$, 
\begin{align}
  \lambda(x)
  &=
  \sqrt{\frac{\pi}{45 G_N}}
  g_*^{1/2}
  m_X \langle \sigma v \rangle 
  \\
  r_{eq}(x)
  &=\frac{Y^{eq}_{\bar{X}}}{Y^{eq}_X}
  =\exp\left(-2\sinh^{-1}\frac{\eta}{2Y_{eq}(x)}\right)
  \\
  Y_{eq}(x)
  &=\sqrt{Y^{eq}_X Y^{eq}_{\bar{X}}}
  \simeq
  \frac{45 g_X}{4\sqrt{2} \pi^{7/2} h_{eff}(x)}
  x^{3/2}\xi^{3/2} e^{-x/\xi}
\end{align}
The values for $g_*^{1/2}, h_{eff}$ are tabulated in
\cite{Gondolo:2004sc}, and
$g_X=2$ is the number of degrees of freedom of $X,\bar{X}$.
$\lambda(x)$ is roughly constant, since the
annihilation cross section is 
\begin{align}
  \langle\sigma v\rangle_{ann}
  &=
  \frac{\pi \alpha_D^2}{m_X^2}
  +
  \frac{\pi (\alpha_D+\alpha_{Z'})^2}{m_X^2}
  \, 
  \equiv
  \frac{\pi \alpha_{\mathrm{eff}}^2}{m_X^2}
  \label{eq:alphaeff}
\end{align}
to leading order in $m_{Z'}/m_X$ (for the couplings we are interested
in, the Sommerfeld enhancement during thermal freezeout is
negligible).

We use an approximation
for $r_\infty$ by integrating the Boltzmann
equation from the freeze-out temperature $x_f$ to today
neglecting $r_{eq}$, and with the boundary condition
$r(x_f) = r_{eq}(x_f)$. 
The value $x_f$ can be found by solving the following
implicit equation,
\begin{align}
  \frac{dr_{eq}(x_f)}{dx}
  &= 
  -\frac{\lambda(x_f) \eta}{x_f^2} r_{eq}(x_f)
\end{align}
with the approximate solution,
\begin{align}
  x_f
  &\simeq
  \xi_f
  \left[\log\left(\xi_f^{3/2} \lambda 
  \frac{45 g_X}{4\sqrt{2} \pi^{7/2} h_{\rm eff}(x_f)}
  \right)
  -\frac12 \log (x_f)
  +\log
  \left(
  1 + \frac16\left(\frac{\eta \lambda}{2x_f^2} \right)^2
  \right)
\right]
  \, .
\end{align}

The value of $r_\infty$ obtained through this procedure is a
good approximation to the numerical solution.
\begin{align}
  r_\infty
  &\simeq
  r_{eq}(x_f)
  \exp
  \left(
  -{\frac{\lambda(x_f) \,\eta}{ x_f}}
  \right)
  \label{eq:rinfty}
\end{align}

Note that $\lambda$ 
is directly related to the relic abundance in the limit
$\eta\to0$ (i.e.~in the absence of an asymmetry),
\begin{align}
  Y_{\infty}^{\eta=0} 
  \simeq
  \frac{x_f}{\lambda(x_f)}
  \,,
\end{align}
so that,
\begin{align}
  r_\infty
  &\simeq
  r_{eq}(x_f)
  \exp
  \left(
  -{\frac{\eta}{Y_{\infty}^{\eta=0}}}
  \right)
\end{align}
For
the parameter space of interest to us, $r_{eq}(x_f) \simeq 1$. 
The dark matter density today is
\begin{align}
  \Omega_X  h^2
  &=
  \frac{\Omega_B h^2}{\eta_B}
  \frac{m_X}{m_p}
  (Y_{X}(x_0) + Y_{\bar{X}}(x_0))
  =
  \Omega_B h^2
  \frac{\eta}{\eta_B}
  \frac{m_X}{m_p}
  \frac{1+r_\infty}{1-r_\infty}
  \label{eq:relic2}
\end{align}
where $\Omega_B, \eta_B$ are the baryonic energy density and
asymmetry respectively, and $x_0$ is the value of $x$ today. 

\begin{figure}[!tp]
  \centering
  \includegraphics[width=0.65\textwidth]{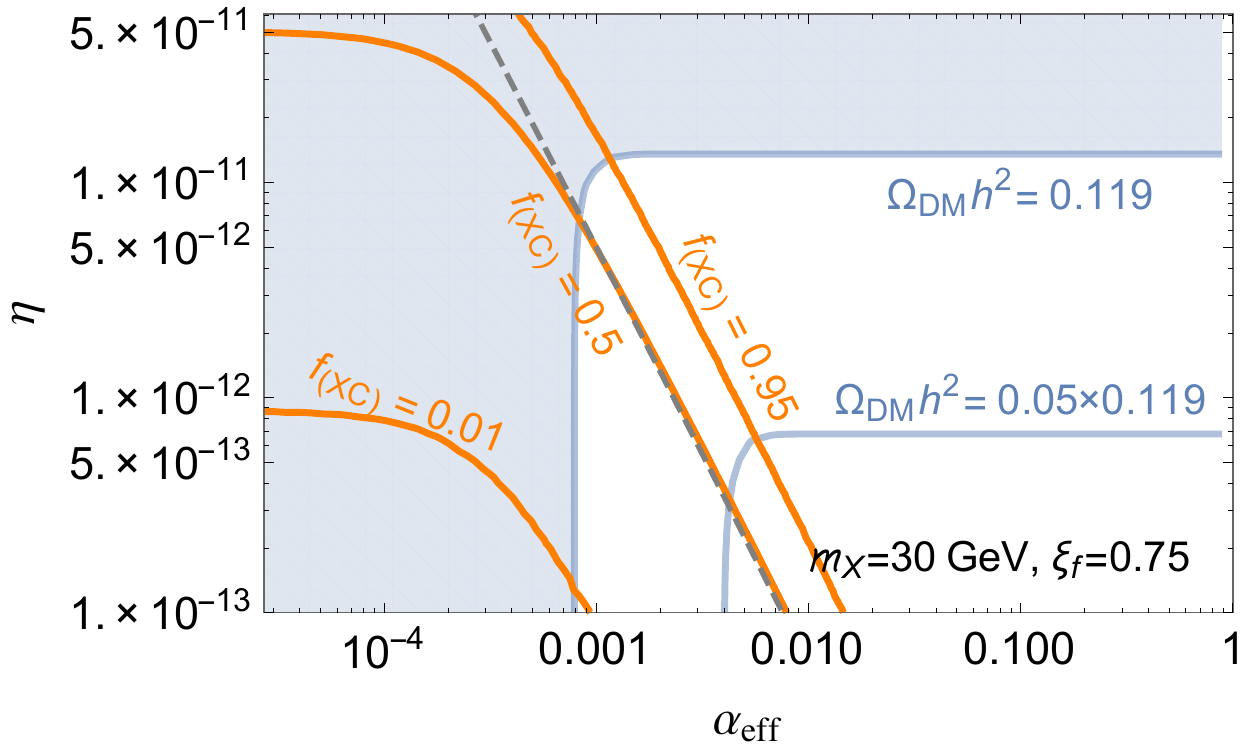}
  \caption{The relic abundance for the $XC$ component of dark matter.
    The effective coupling $\alpha_{\mathrm{eff}}$ is defined in
    equation~\ref{eq:alphaeff}.
    The blue  lines are contours of $\Omega_{X}h^2$ and
    the orange lines are contours of $f_{(XC)}$. The gray dashed line
    shows the turn-over point where $\eta = Y_\infty^{\eta=0}$. We
    chose $m_X=30$ GeV
  for this plot with $\xi_f=0.75$.}
  \label{fig:asymm-symm}
\end{figure}

In figure~\ref{fig:asymm-symm} we show the region of
$\alpha_{\mathrm{eff}}$--$\eta$ parameter space where a significant
symmetric dark
matter relic abundance survives in the presence of an asymmetry. 
The  blue curves are contours of constant dark matter relic density
$\Omega_X$, and the orange lines denote the value of $f_{(XC)}$,
defined in equation~\ref{eq:asymm}.
The shaded region is excluded in this model as too much
dark matter would survive.

We see that there are two limiting behaviors of the solutions. When
$\eta\ll Y_{\infty}^{\eta=0}$, the result is essentially the symmetric
freezeout value. Conversely, for $\eta > Y_{\infty}^{\eta=0}$, the
relic abundance is mainly set by the asymmetry. For appreciable
amounts of both
symmetric and asymmetric components, we choose parameters in the
turnaround region, $\eta\sim Y_{\infty}^{\eta=0}$.

\subsection{Cooling} 

We now  consider how dark matter
haloes might cool through dissipation to  form compact objects with
enhanced density. The cooling calculation and requisite parameter
space follows closely the one outlined in~\cite{Fan:2013yva}, with the
modification due to additional $\bar{X}$
particles~\cite{Agrawal:2017rvu}.
                   
The dissipative component of dark matter behaves in analogy to baryonic
matter during galaxy formation. Clumps of $X,C,\bar{X}$ accrete onto
the CDM halo, and are subsequently shock heated to its virial temperature. If this temperature is larger than the
binding energy of the $X C$ bound state, then the charged $X$ and $C$  
can radiate. The lighter $C$ dominates the cooling and leads to collapse into a 
region with lower velocity than that in the dark matter halo and can in principle form a disk. The
cooling time of dissipative dark matter must be smaller than the age
of the universe (more precisely the dynamical time of the virialized
halo) for disks or compact structures to form.

The subsequent behavior of the dark matter gas depends on the nature
of the collapse and other details of the model. For baryonic matter
this collapse leads to sites of star formation. In  principle high
density compact objects can result from this collapse in the dark
sector as well, though in the absence of additional forces one would not expect nuclear burning or feedback.

\begin{figure}[!t]
  \centering
 \includegraphics[width=0.65\textwidth]{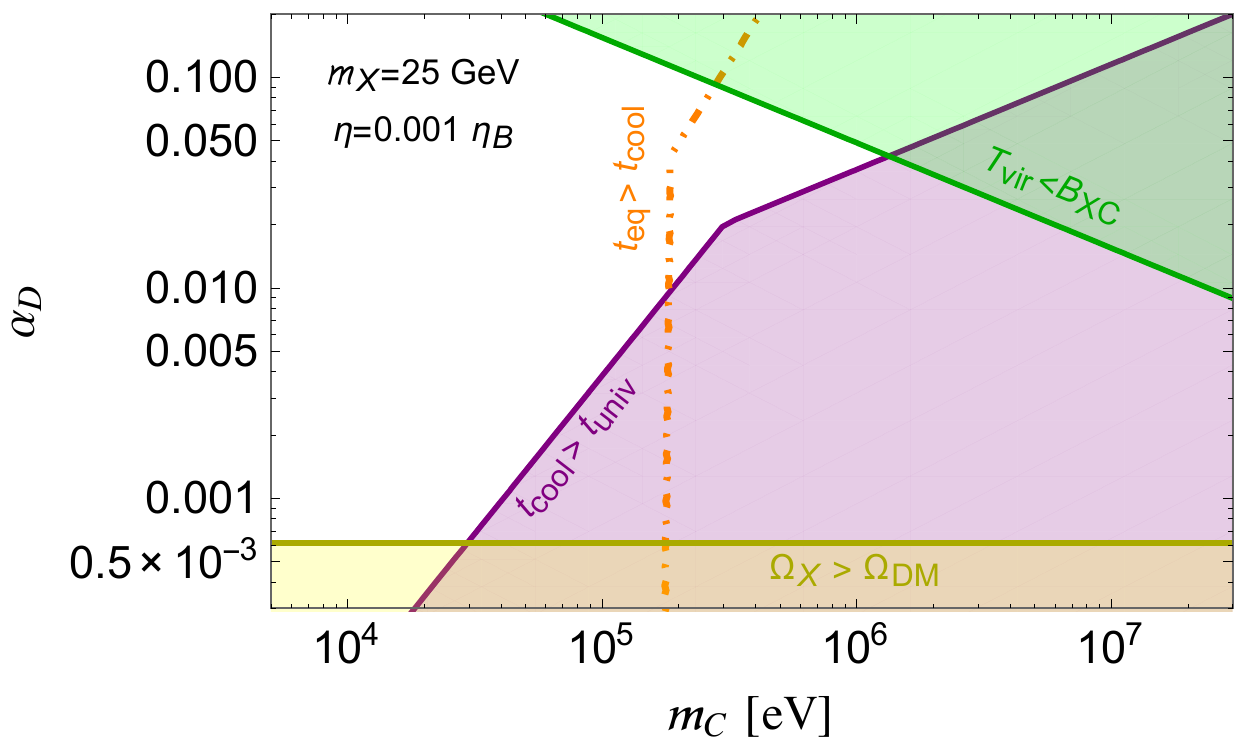}
  \caption{
    Constraints on the XC model in the $m_C$--$\alpha_D$ plane. Shown
    are constraints from relic abundance (yellow), reionization in
    galaxies (green),
    cooling rate (purple), and region where the $XC$ particles are in
    thermal
    equilibrium (orange dot-dashed). We have ignored the contribution
    of $\alpha_{Z'}$ to the relic abundance for this plot.
}
  \label{fig:cooling2}
\end{figure}

We assume the initial condition of a shock heated gas
inside a CDM halo of mass $M_{gal}\simeq10^{12}
M_{\odot}$ and virial radius $R_{vir}\simeq 110 \kpc$.
We estimate the cooling timescale for the scenario where the $XC$ dark
matter initially falls into the CDM halo with a Navarro-Frenk-White
(NFW) profile with a scale radius
$R_s = 20\kpc$. The virial temperature for this case is
\begin{align}
  T_{vir}
  &=
  \frac{G_N M_{s} \mu}{5 R_{s}}
  \, ,
\end{align}
where $\mu = m_X/(1+f_{(XC)})$ is the average mass of a particle in
the dark matter gas, and $M_s \simeq 2\times10^{11} M_\odot$ is total
halo mass inside the scale radius $R_s$.
The density of $X$,$\bar{X}$ and $C$
particles inside the scale radius is 
\begin{align}
  n_X + n_{\bar{X}}
  &=
  f
  \frac{1}{m_X}
  \frac{M_{s}}{\frac43 \pi R_s^3}
  \\
  n_C
  &=
  f_{(XC)} (n_X + n_{\bar{X}})
\end{align}
where we have expressed
the density of $C$ particles in terms of the asymmetry
$f_{(XC)}$ defined in equation~\ref{eq:asymm}.

We assume
Bremsstrahlung and Compton
scattering off the dark CMB photons that can lead to
cooling~\cite{Fan:2013yva} (for a detailed analysis of cooling in
the dark sector
see~\cite{Rosenberg:2017qia}). 
The time scale for bremsstrahlung is,
\begin{align}
  t_{brem}
  &=
  \frac{3}{16}
  \frac{n_X+n_{\bar{X}}+n_C}{(n_X+n_{\bar{X}}) n_C}
  \frac{m_C^{3/2} T_{vir}^{1/2}}{\alpha_D^3}
    \label{eq:cool-brem}
\end{align}
The Compton cooling off dark CMB occurs on a timescale 
\begin{align}
  t_{comp}
  &=
  \frac{135}{64\pi^3}
  \frac{n_X + n_{\bar{X}} + n_C}{n_C}
  \frac{m_C^3}{\alpha_D^2 (T_D^0(1+z))^4}
  \label{eq:cool-compton}
\end{align}
The time scale for equilibration between $X$ and $C$ is
\begin{align}
  t_{eq}
  &=
  \frac{m_X m_C}{2 \sqrt{3\pi} \alpha_D^2}
  \frac{(E_C/m_C)^{3/2}}
  {n_C \log \left(1+\frac{v_C^4 m_C^2}{\alpha_D^2 n_C^{2/3}}\right)}
\end{align}If the
$C$ particles are in kinetic equilibrium with $X$ and $\bar{X}$, i.e.\
the equilibration timescale is shorter than the cooling time scale,
the cooled compact objects will be composed of all of $X, \bar{X}, C$
particles. On the other hand, if $t_{eq} > t_{cool}$, then the $C$
particles cool but cannot cool $X$ and $\bar{X}$ particles through
equilibrium processes.

\subsection{Direct detection}
In principle, the kinetic mixing of the $Z'$ with the photon in the
portal models can give rise  direct detection, which in principle
could be a significant constraint.  
The direct detection constraints for millicharged dark matter
(i.e.~through the massless $Z'$ portal) are very strong, with the
WIMP-nucleon cross section estimated by
\begin{align}
  \sigma^{(n)}
  &=
  \frac{Z^2}{A^2}
  \frac{16\pi \alpha^2\epsilon^2  \mu_n^2}{q^4}
  \sim
  10^{-45}\cm^2
  \left(\frac{\epsilon}{10^{-10}}\right)^2
  \left(\frac{30\MeV}{q}\right)^4
  \,.
\end{align}
where $q$ is the typical momentum transfer and $Z$ and $A$ are the
atomic and mass number of the target.
However, in much of the parameter space, the millicharged dark matter
is evacuated from the disk by shock waves generated during supernova
explosions, and is shielded from (re)entering the disk 
by the large-scale magnetic field of the
galaxy~\cite{Chuzhoy:2008zy,McDermott:2010pa}. This mechanism is
efficient for 
\begin{align}
  5.4\times 10^{-13} \frac{m_X}{\GeV}
  < \epsilon <
  3.4 \times{10^{-4}}\sqrt{\frac{m_X}{\GeV}}
\end{align}
which is a very large range of $\epsilon$ for $m_X \sim 30 \GeV$.

The constraints for the massive $Z'$ portal depend on the mass of the
$Z'$. For our case, where $m_{Z'}\gg q$, the direct detection cross
section is,
\begin{align}
  \sigma^{(n)}
  &=
  \frac{Z^2}{A^2}
  \frac{16\pi \alpha^2\epsilon^2  \mu_n^2}{m_{Z'}^4}
  \sim
  10^{-45}\cm^2
  \left(\frac{\epsilon}{10^{-5}}\right)^2
  \left(\frac{10\GeV}{m_{Z'}}\right)^4
  \,.
\end{align}

In fact this constraint is further relaxed in our setup.  The
$X,\bar{X}$ component of dark matter makes up only a fraction of the
total dark matter. Most of $X$ and $\bar{X}$ are in collapsed objects,
so the direct detection signal exists only if these objects overlap
with our solar system. Finally, the velocity dispersion in the
collapsed object is much lower than the velocity dispersion for the
CDM halo, leading to typical recoil energies lower than the
experimental threshold for most direct detection experiments.  We
conclude that direct detection does not put strong model-independent
constraints on the model we are investigating.

\subsection{Constraints on compact objects}
We  now estimate the size of compact objects in our models and
evaluate existing constraints. We show these constraints
in figure~\ref{fig:survival-zp}.
\begin{itemize}
  \item {\bf Total number}\\
Given a certain mass of the objects and the fraction of dark matter in
the $XC$ component, we can derive a bound on the number of such
objects in our halo,
\begin{align}
  \mathcal{N}_{total}
  \lesssim
  10^6
  \left(\frac{f}{0.01}\right)
  \left(\frac{10^4 M_\odot}{M}\right)
  \left(\frac{M_{DM,total}}{10^{12} M_\odot}\right)
  \, .
  \label{eq:total}
\end{align}
in the entire galaxy.
If this bound is not saturated the
rest of dissipative dark matter might be in more diffuse clouds.

\item {\bf Longevity} \\
The symmetric component in dense objects can annihilate, putting a
constraint on the size of a dark photon gauge coupling for a given value
of $M$ and $R$. This constraint is
relevant for the existence of compact objects even if there is no
$\gamma$-ray signal.
\begin{align}
  n_{X} \sigma_{X\bar{X}} v 
  &=
  \frac{1+f_{(XC)}}{2}
  \frac{M}
  {m_X \left(\frac43 \pi R^3\right)}
  \left(
  \frac{\pi \alpha_D^2}{m_X^2}
  +\frac{\pi (\alpha_D+\alpha_{Z'})^2}{m_X^2}
  \right)
  S_{ann} (\alpha_D / c_{XC})
  <
  H_0
\end{align}
where 
we have ignored corrections of
$\mathcal{O}(m_{Z'}/m_X)$, and ${S}_{ann}$ is the
Sommerfeld
enhancement, 
\begin{align}
  S_{ann} (\zeta)
  &=
  \frac{2 \pi \zeta}{1-e^{-2\pi\zeta}}
  \,.
\end{align}

\item {\bf Further cooling}\\

If the virial velocity for a compact object is higher than the binding
energy for the $XC$ system, then there may be a non-negligible ionized
$C$ component, which can lead to further cooling and collapse. 
For a virially supported object, we can impose the following
conservative bound,
\begin{align}
  T_{vir} = \frac{ G_N M \mu}{5R} < \frac12 m_C \alpha_D^2
  \,.
\end{align}
We impose this constraint for our analysis, but in principle such
objects can also be stabilized by other mechanisms.
We are also ignoring other cooling precesses such as atomic and molecular
cooling which could cool the compact object further.

\item {\bf Tidal Stripping}\\
The tidal radius for a compact object of mass $M$ at a distance $r$
from the galactic center is,
\begin{align}
  R_{tidal}
  &\simeq
  r
  \left(\frac{M}{3M_{enc}(r)}\right)^\frac13
  \,.
\end{align}
Objects larger than this radius will be tidally disrupted. If the density
profile of the compact object is constant, then  objects of
all size with the given density would not remain. But if the density
towards the inner parts the compact object is higher, then tidal
disruption can leave behind a dense core of the original object.

In the inner galaxy, the mass enclosed is dominated by baryons in the
form of the bulge and the stellar disk. We use the parametrization of
the bulge and disk from~\cite{Binney:1996sv,McGaugh:2008nc},
\begin{align}
  \rho_B(b)
  &=
  \frac{\rho_{B,0}}{\eta \zeta b_m^3}
  \frac{e^{-(b/b_m)^2}}{(1+b/b_0)^{1.8}}
\end{align}
with $b^2 = x^2 + (y/\eta)^2 + (z/\zeta)^2$,
$\rho_{B,0} = 3.5\times10^{12} M_\odot, b_m = 1.9 \kpc, b_0 =
0.1 \kpc, \eta = 0.5$ and $\zeta=0.6$.

The stellar disk is modeled as
\begin{align}
  \Sigma_\star (r)
  &=
  \Sigma_0 e^{-r/r_d}
\end{align}
with $\Sigma_0 = 562 M_\odot /\pc^2$ and scale radius $r_d = 3 \kpc$.
As a benchmark, we impose the constraint that the compact object is
smaller than the tidal radius at $\sim$ 0.3 kpc from the galactic center,
\begin{align}
  R 
  &\lesssim
  (0.3\kpc)
  \left(\frac{M}{3M_{enc}(0.3\kpc)}\right)^\frac13
  \,,
  \\
  M_{enc} (0.3\kpc) 
  &\simeq
  \int_0^{0.3\kpc} dx\, dy\, dz\, \rho_B(b)
  +\int _0^{0.3\kpc} 2\pi r' \Sigma_\star (r')dr'
  =3\times10^{8} M_\odot
  \,.
\end{align}

\end{itemize}

\section{The Galactic Center Excess}
\label{sec:GCE}
A number of analyses have found that the FERMI telescope has observed an excess of gamma ray emission from
the center of the
galaxy~\cite{Goodenough:2009gk,
Hooper:2010mq,Daylan:2014rsa,Calore:2014xka,TheFermi-LAT:2015kwa}.
There are also
hints for excess gamma-ray emission from
Andromeda~\cite{Ackermann:2017nya} and from the dwarf
galaxy Reticulum II ~\cite{Geringer-Sameth:2015lua,Bonnivard:2015tta,Fermi-LAT:2016uux}.
The origin of the GC excess is intensely debated, but one exciting
possibility is that it originates from 
dark matter
annihilations~\cite{Goodenough:2009gk,Hooper:2010mq,
Daylan:2014rsa,
  Alves:2014yha,Berlin:2014tja,
Agrawal:2014una,Agrawal:2014aoa,
Izaguirre:2014vva,Ipek:2014gua,Boehm:2014bia,
Abdullah:2014lla,Martin:2014sxa,Bell:2014xta,Freytsis:2014sua,
Agrawal:2014oha,
Buckley:2014fba,Elor:2015tva,Kaplinghat:2015gha,Karwin:2016tsw}.

However, Refs.~\cite{Lee:2015fea,Bartels:2015aea} showed that there are some hints 
for a statistical preference for point-source emission in
the GCE (but this conclusion might be sensitive to structure in the
diffuse background that is not modeled by the range of background
models considered, see also~\cite{Horiuchi:2016zwu}). The morphology
of the excess if pointlike would be in tension with the smooth distribution
expected for a CDM
annihilation~\cite{Clark:2016mbb,Fermi-LAT:2017yoi}. 
Most physicists would then attribute
the observed point-like spectrum to a new population of
Millisecond Pulsars (MSPs)~\cite{Brandt:2015ula,Fermi-LAT:2017yoi}. We
investigate the alternative possibility that these point sources
could arise from dark matter annihilation in compact objects.

We highlight the parameters of our model that can reproduce this point
source signal, specifically the mass $m_X$ and the coupling
$\alpha_{Z'}$, and the size of objects, $\{M,R\}$. 
We choose the simplified limit where all objects have similar masses
and sizes, and have a uniform density.
It would be straightforward to vary these assumptions to include any
other mass function and a realistic density profile and distribution that
might be motivated by N-body simulations. If there is a very sizeable
spread in the mass function, there will be a large number of brighter
compact objects, which should have been resolved as point sources by
Fermi.
We assume the spatial distribution of the compact objects in the inner
galaxy to be spherically
symmetric with an NFW-squared radial
profile (more precisely $r^{-2.5}$) in the inner galaxy. This is
motivated by the observed
morphology of the galactic center excess.

The mass $m_X$ can
be fixed by finding the best-fit spectrum to the excess.  The point
source analysis in~\cite{Lee:2015fea} uses a single wide energy bin,
which is unsuitable for
a spectral analysis. Instead we find the best-fit spectrum following the
analysis in~\cite{Calore:2014xka,Agrawal:2014oha}.  We then consider
the annihilation rate and the size and mass of compact objects
required to reproduce the observed point-source excess.

\subsection{Spectrum for the Galactic Center excess}
\label{app:GCAnalysis}
\begin{figure}[t]
  \centering
  \includegraphics[width=0.45\textwidth]{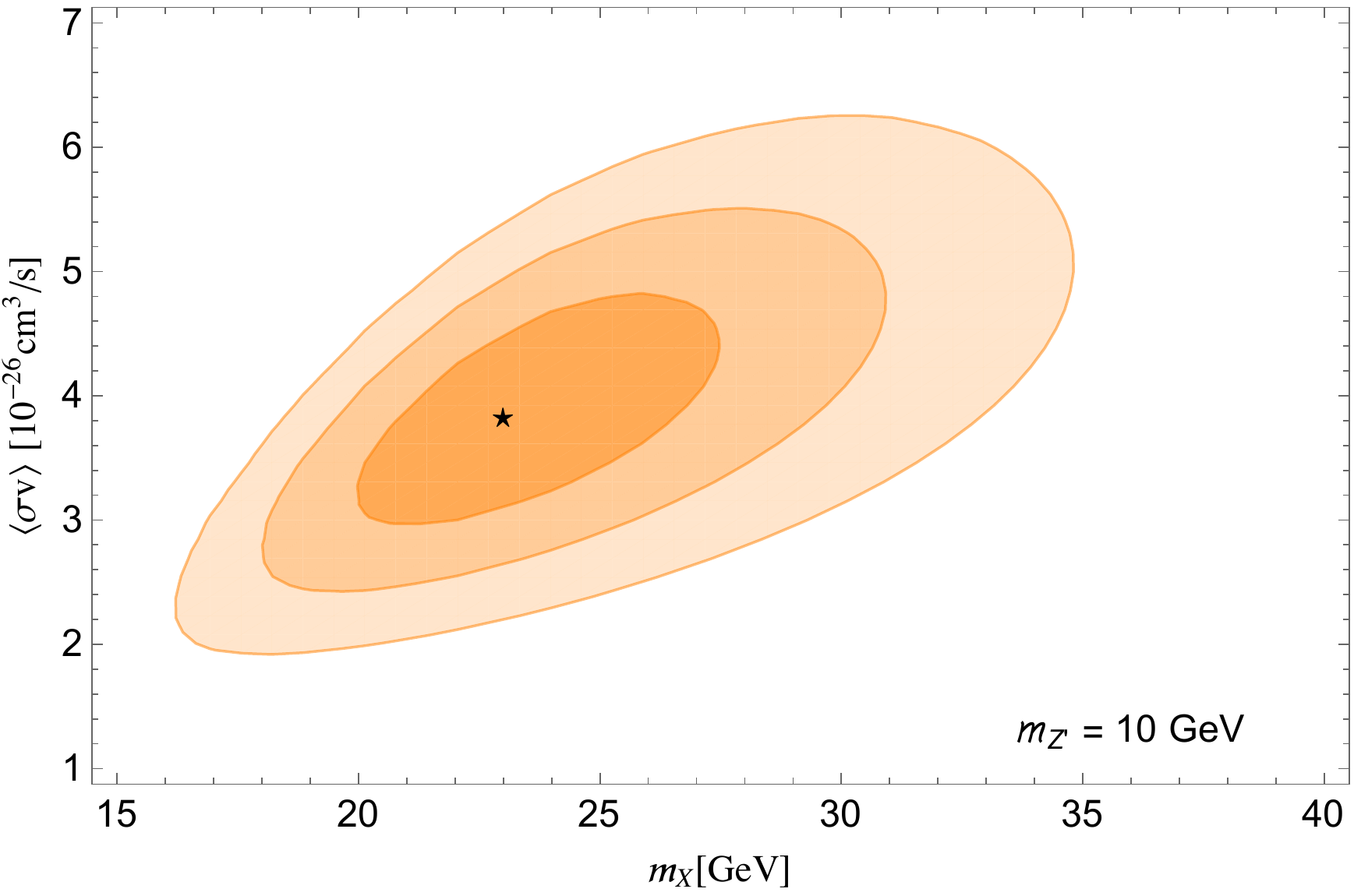}
  \qquad
  \includegraphics[width=0.45\textwidth]{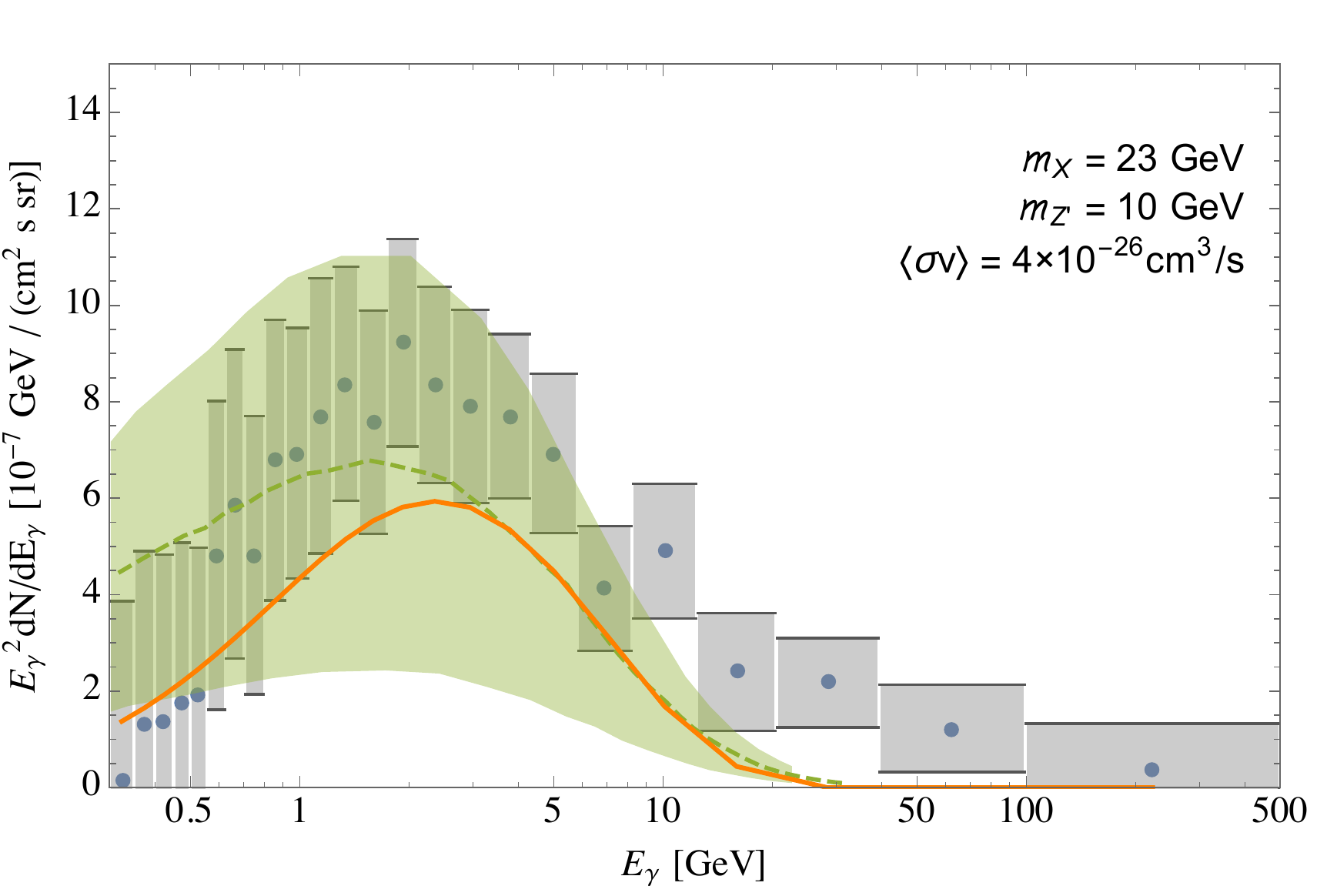}
  \caption{Left: We show $1,2,3\sigma$ contours around the best-fit
    spectrum for the
  $Z'$ model. Right: The spectrum of photons obtained for the best-fit
  point in our model (orange) overlaid with the spectrum for the
  excess from Calore et
  al~\cite{Calore:2014xka} (gray) and recent Fermi
  results~\cite{Fermi-LAT:2017yoi} (green). Note that the gray error bands are
  highly correlated, and the orange curve does fit the data reasonably
  well. For this plot we assume an NFW profile with a region of
    interest in~\cite{Calore:2014xka,Agrawal:2014oha} with 
    $J = 2.0\times10^{23} \GeV^2/\cm^5$.
}
  \label{fig:gce}
\end{figure}

For the spectral fit in this section, we will follow the analysis
of~\cite{Calore:2014xka}. The region of interest (ROI) in this
analysis extended to a $\pm20^\circ$ square around the galactic
center, with the inner $2^\circ$ latitude masked out. We further
choose an NFW profile ($\rho_\odot = 0.4 \GeV/\cm^3 , R_s = 20 \kpc,
\gamma = 1.20$), which translates to $J=2.0\times10^{23} \GeV^2 /
\cm^5$.

The flux from DM annihilation with DM mass $m_X$ is (for Dirac
fermions)
\begin{align}
  \frac{d\Phi}{dE}
  &=
  \frac{J \langle \sigma v\rangle}{16\pi m_X^2}
  \left< \frac{dN}{dE} \right>
\end{align}
where the flux factor $J$ is the line-of-sight integral of the dark
matter density, and $dN/dE$ is the spectrum of photons per
annihilation. Given the $J$-factor and the spectrum per annihilation
for different masses,
we can in principle find the best-fit $m_X$ and $\langle \sigma v
\rangle$. We will
use only the $m_X$ value found this way, and our best-fit $\langle \sigma
v \rangle$ will instead be determined by fitting to the point source
flux in the next section.

When $\alpha_D > \alpha_{Z'}$,
the dominant annihilation mode that produces photons 
is $X\bar{X} \to \gamma_D Z'$. The $Z'$ then
decays to SM fermions through the $\epsilon$ mixing which produces
a continuum photon spectrum. 
The spectrum  can be
calculated by boosting the  photons in the rest frame of
the $Z'$ to the galactic frame. The spectrum of photons produced in
such cascade decays was also considered in a model-independent
analysis in~\cite{Elor:2015bho}.

The spectrum of photons in the rest frame of $Z'$ 
is
\begin{align}
  \left.\frac{dN(E)}{d E} \right|_{Z'}
  &=\sum_f br(Z'\to f)
  \left.\frac{dN_f(E)}{d E} \right|_{Z'}
\end{align}
where the sum is over the spectrum from each of the decay modes of the
$Z'$ through the kinetic mixing~\cite{Curtin:2014cca}, and
$\frac{dN_f(E)}{d E}$ is the spectrum of photons from
decay  $Z'\to f$. 
The spectrum from various standard model final states can be simulated
using {\tt PYTHIA}~\cite{Sjostrand:2006za,Sjostrand:2007gs}, and has
been tabulated in
{\tt PPPC 4 DM ID}~\cite{Cirelli:2010xx,Ciafaloni:2010ti}. 

The spectrum of photons per annihilation, in the galactic frame, is
obtained by convolving the spectrum in the rest frame of $Z'$
with a unit normalized ``box''~\cite{Agrawal:2014oha}
\begin{align}
  \label{eq:boosted}
  \left< \frac{dN(E) }{dE}\right >
  &=
  \frac{1}{x_+-x_-} \int_{Ex_-}^{Ex_+} \frac{d E' }{E'}
  \left.\frac{dN(E')}{d E'} \right|_{Z'}
\,,
\end{align}
and
\begin{align}
  \label{eq:Epm}
  x_\pm= 
  \frac{4m_X^2 + m_{Z'}^2}{4 m_X m_{Z'}}
  \left(
  1\pm 
  \frac{4m_X^2 - m_{Z'}^2}{4m_X^2 + m_{Z}^2}
  \right)
\end{align}

In figure~\ref{fig:gce} we show the $\chi^2$-contours for the photon
spectrum fit to the observed excess, as a function of the dark matter
mass and the cross section (given the astrophysical
$J$-factor) and the spectrum of the best-fit point. The
best fit dark matter mass for our
model, $m_X\simeq 25\GeV$ is seen to provide a good fit to the
observed spectrum of the excess.

\subsection{The Point-like GeV Excess from  Compact Objects
in Dissipative Dark Matter}

According to
 ~\cite{Lee:2015fea}
(see also~\cite{Bartels:2015aea,Fermi-LAT:2017yoi}), the signal
appears to originate
from point-like sources that are distributed with an
NFW-squared profile. The region of interest for this analysis was
taken to be inner $30^\circ$ with $|b|<2^\circ$ masked.
The pixel size for the analyses is $0.5^\circ$ to a side, which
translates to about $75\pc$ at the center of the galaxy. The point
spread function varies between $0.05^\circ$ to $0.2^\circ$ for the
energy range $1$--$10\GeV$. Therefore,
any objects smaller than about $10-100\pc$ would give rise to 
non-Poissonian photon statistics, appearing as point sources. 
We saw in
section~\ref{sec:fragments}
that for our chosen
parameters we expect our compact objects to be $\mathcal{O}$(10--100)$\pc$ in
size, and hence to appear as point sources.

The analysis predicts a certain number of objects
in the inner galaxy to account for the total flux absorbed by the
point source template. 
In the inner $10^\circ$ of 
the galaxy with $|b|\geq 2^\circ$, $86^{+32}_{-25}$ point sources
can explain about half of the excess from the galactic center when
the Fermi 3FGL point sources are masked. For the unmasked analysis
in~\cite{Lee:2015fea} the corresponding number is $132^{+31}_{-25}$.

We note that the flux and the number of objects is roughly fixed
independently in addition to being constrained by the total amount of
flux. If we had a much larger number of objects (with a corresponding
smaller flux per object), they would have photon-per-pixel statistics
closer to a smooth morphology predicted by standard CDM
substructure.  Much fewer objects would need to be brighter to
account for the total flux, and hence would be above the threshold for
detection as point sources by Fermi. 

We next turn to the normalization of our signal, which we fit
to~\cite{Lee:2015fea}.
The normalization depends on the $J$-factor, on the total annihilation
cross section $\langle \sigma v
\rangle$ and on the number of photons per annihilation. 
\begin{align}
  \Phi
  &=
  \frac{\langle \sigma v\rangle J}{16\pi m_X^2}
  \langle N_\gamma \rangle
  \\
  \langle N_\gamma \rangle
  &\equiv
  \int_{E_{min}}^{E_{max}} dE 
  \left<
  \frac{dN}{dE}
  \right>
\end{align}
The total annihilation rate $X\bar{X}\to Z'\gamma_D$ is 
\begin{align}
  \langle \sigma v \rangle
  &=
  \frac{\pi \alpha_{Z'}\alpha_D}{m_X^2}
  S_{ann} (\alpha_D / c_{XC})
\end{align}
to leading order in $m_{Z'}/m_X$. 
The velocity of dark matter in the compact object ($c_{XC}$) is given in
equation~\ref{eq:dmvel}.

In this analysis, a single energy bin from 1.893--11.943 GeV was
used. For $m_X = 25\GeV$, $m_{Z'}=10\GeV$, this corresponds to,
\begin{align}
  \langle N_\gamma \rangle
  \simeq
  1.8
\end{align}

The $J$-factor associated with the point sources is given by,
\begin{align}
  J
  &=
  \sum_i^{\mathcal{N}} J_i
  =\mathcal{N}_{src} \bar{J}
\end{align}
where $\bar{J}$ is the $J$-factor for each point source if they are
all identical. In the analysis of~\cite{Lee:2015fea}, the flux from
each source was estimated to be
$\bar{\Phi}=1.4\pm0.3\times10^{-10}$ photons/cm$^2$/s. 
The $J$-factor for each point source is
\begin{align}
  \bar{J}
  &=
  \left( 1 - f_{(XC)}^2 \right)
  \left(\frac{M^2}{\frac43 \pi R^3} \right)
  \frac{1}{r_\odot^2}
\end{align}
where $f_{(XC)}$ is the fraction of the dissipative component in bound
states. 
The flux from each point source is,
\begin{align}
  \bar{\Phi}
  &=
  \frac{\langle \sigma v\rangle \bar{J}}{16\pi m_X^2}
  \langle N_\gamma \rangle
  \\
  &\simeq
  1\times10^{-10} \frac{1}{\cm^2 \s}
  \left( 1 - f_{(XC)}^2 \right)
  \left(\frac{10\pc}{R} \right)^3
  \left(\frac{M }{10^4 M_\odot}\right)^2
  \left(\frac{25\GeV}{m_X}\right)^2
  \left(\frac{\langle \sigma v\rangle}{1.4\times10^{-24}
\cm^3/\s}\right)
\left(\frac{\langle N_\gamma \rangle}{1.8}\right)
\end{align}
where the numerical prefactor above is the consistent with the typical flux
from a point
source in the analysis of~\cite{Lee:2015fea,Lee:2014mza}.
The value of $\langle\sigma v\rangle$ is enhanced by the Sommerfeld
effect, and therefore can be much larger today than the canonical thermal
freezeout 
value of $2.2\times 10^{-26}\cm^3/\s$ without affecting the relic
abundance calculation.

\begin{figure}[!tp]
  \centering
  \includegraphics[width=0.65\textwidth]{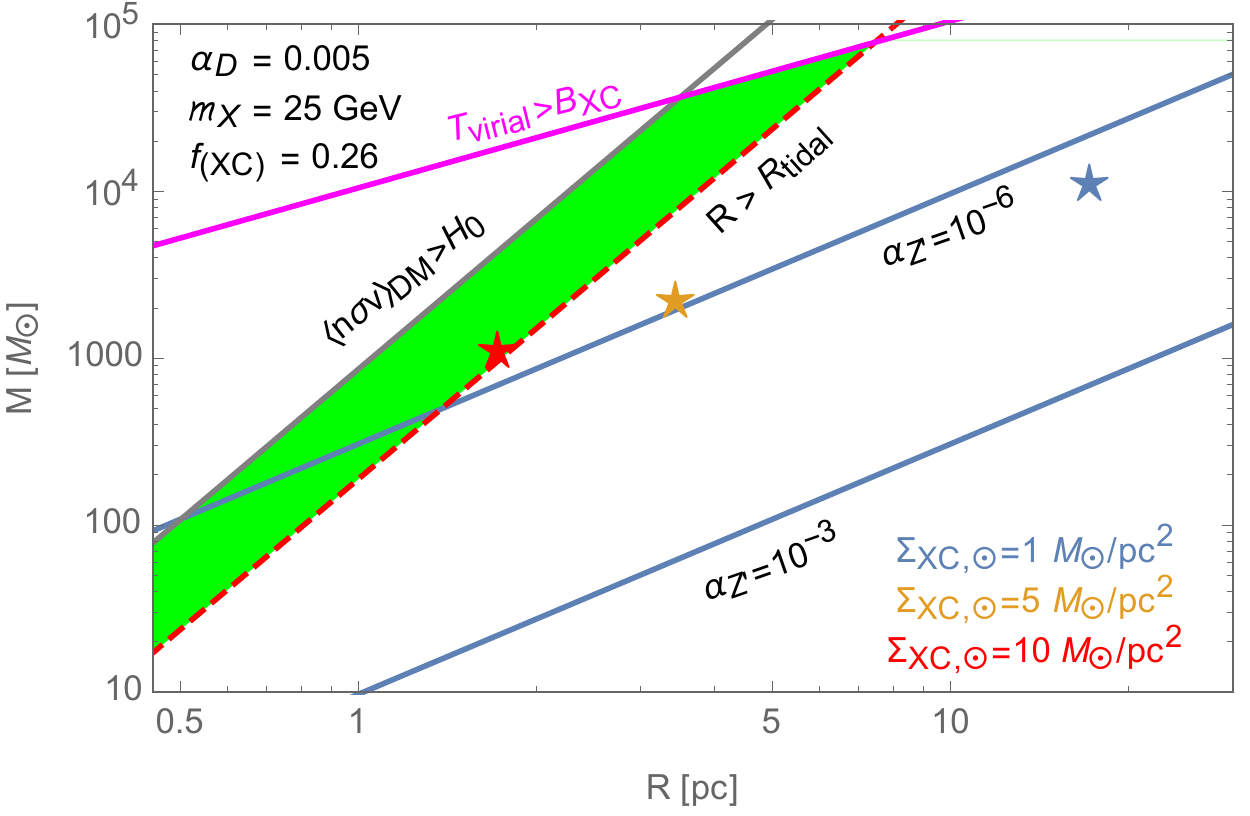}
  \caption
  { 
    We show constraints on compact objects
    as a function of their mass $M$ and radius $R$. The green shaded
    region is the preferred region for our analysis.
    In the region above the gray curve
   $\bar{X}$ annihilates away within the
    lifetime of the universe. 
    The region above the 
    magenta line has $T_{virial} > B_{XC}$, so that the compact object would be
    unstable to further cooling. Objects to the right of the red
    dashed line can be tidally disrupted in the inner galaxy.
    The blue
    lines are contours of $\alpha_{Z'}$ to obtain the galactic center
    excess flux from $X\bar{X} \to Z'\gamma_D$ annihilation.  
    We indicate a few benchmark values of $\{M,R\}$ (stars) which are
    estimates of the mass and size of compact objects we get from
    fragmentation for a given dark disk column density.
  }
    \label{fig:survival-zp}
  \end{figure}

In figure~\ref{fig:survival-zp} we see the parameter space where the
dark matter annihilations can explain the GCE. We also show
constraints on the parameter space where the compact objects can
undergo tidal disruption, annihilate away within the age of the
universe, or are unstable to further cooling. We also show benchmark
points from our instability analysis in section~\ref{sec:compact-objects} for
a few choices of the local dark disk density.

As noted above, we need an $\mathcal{O}(100)$ compact objects of mass
$10^4 M_\odot$ each to account for the total flux of the GCE. For a
CDM component with an NFW profile the 
total mass of dark matter in the inner galaxy region is $\sim10^9
M_\odot$. Therefore, $XC$ dark matter can be as little as $10^{-3}$ of
the total dark matter in the inner galaxy while still giving us interesting
signals. Intriguingly, the benchmark values from our stability
analysis, $\Sigma_{XC,\odot}\sim$ 1--10 $M_\odot/\pc^2$,
translate into a total dark disk mass of $10^9$--$10^{10} M_\odot$,
or $10^{-3}$--$10^{-2}$ fraction of the total galactic dark matter.

\section{Conclusion}
\label{sec:conclusion}
Dissipative dark matter models can give rise to clumps of dark
matter with enhanced density. Even if dissipative dark
matter makes up a tiny fraction of the total dark matter density,
these clumps can be the dominant site of annihilation, leading to
novel indirect detection signatures.

We have considered a simple example of such a dissipative dark matter
model and estimated the size and mass of collapsed objects we might
expect as a function of the model parameters. We have shown that these
objects could be small enough to
appear as point sources in Fermi-LAT. The spectrum depends on the
portal connecting the dark sector with the Standard Model. 

With an additional massless dark photon, dark matter particles can
annihilate to  photon and a dark photon, which appear as a line
spectrum in Fermi-LAT. A line from point sources would be a smoking
gun signal of dissipative dark matter models.
If the portal is a massive vector, the annihilation then leads to
continuum photons, with a spectrum of that 
resembles the spectrum of photons from the $b\bar{b}$ final state.
This can potentially provide an explanation of the
point-source origin of the galactic center
excess as arising from dark matter annihilation in these objects. 

The point source morphology and continuum spectrum of gamma rays are
characteristic signals of millisecond pulsars but as we have
shown might possibly occur in certain classes of dark matter models.
Our analysis gives us a target which we can try and distinguish from
MSPs.

The authors of~\cite{Calore:2015bsx} consider observations of pulsars
in other wavelengths and conclude that pulsar surveys in the radio
frequencies will potentially be able to detect millisecond pulsars in
the bulge.  The absence of any signal in other wavelengths will
strength to the hypothesis of dark matter compact objects as the
origin of this signal. 

The angular size of dark matter clouds in the inner galaxy can be
comparable to the resolution of Fermi. If there are larger clouds, or
clouds closer to us, Fermi-LAT might be able to resolve the structure,
definitively discriminating it from pulsars.

UCMHs that form from large fluctuations in the power spectrum
have also been considered as potential dark point sources~\cite{
Berezinsky:2003vn,Ricotti:2009bs,Scott:2009tu,Josan:2010vn,Erickcek:2011us,Bringmann:2011ut,Berezinsky:2013fxa,Clark:2015sha,Aslanyan:2015hmi,Clark:2016pgn}.
For the UCMH, in addition to annihilation of dark matter
within the UCMH, there will be annihilations of dark matter in the
UCMH and in halo, leading to a different radial
dependence of the excess. There will also be annihilations in the halo
with a diffuse morphology. These features will help distinguish UCMHs
from dissipative compact objects. Further, due to dissipation the
compact objects considered in this paper will be concentrated towards the
center of the halo (possibly along a plane), whereas 
we expect UCMHs to be isotropically distributed throughout the halo.

Very little is known about dark matter, and even less about components
which might be a small fraction. However, it is precisely these
small but interesting components -- not unlike baryons
-- which might provide the most spectacular signals from dark matter.

\acknowledgments
We thank Mariangela Lisanti for very useful comments on an earlier
version of this manuscript. We are grateful to Francis-Yan
Cyr-Racine, Matt Reece and Jakub Scholtz for discussions and comments
on the manuscript.
This work is supported by NSF grants PHY-0855591 and PHY-1216270.

\bibliographystyle{utphys}
\bibliography{ref.bib}
\end{document}